%\magnification=\magstep3
\magnification=1200
%\magnification=1000
%\magnification=\magstep1
%\vsize=20truecm
%\voffset=1.00truein
\settabs 18 \columns
%\hoffset=3.75truecm
%\hoffset=1.00truein
%\hsize=14truecm

\input epsf

%\nopagenumbers
%\baselineskip=17 pt
%\baselineskip=12 pt
\baselineskip=15 pt
%\ifnum\pageno=1
\topinsert \vskip 3.50cm
\endinsert
%\vsize=7.5in
%\fi

%\def\mybox{\sqcap\kern-.66em\sqcup\kern.65em}
\def\sqr#1#2{{\vcenter{\vbox{\hrule height.#2pt
 \hbox{\vrule width.#2pt height#1pt \kern#1pt
 \vrule width.#2pt} \hrule height.#2pt}}}}

\def\operp{\hbox{${\kern+.25em{\bigcirc}
\kern-.85em\bot\kern+.85em\kern-.25em}$}}
%\def\gapprox
%{\hbox{$
%\smash{lower0.5ex\hbox{$\scriptstyle>$}} \atop
%\smash{raise0.3ex\hbox{$\scriptstyle \sim$}}
%$}}
%\def\lapprox
%{\hbox{$
%\smash{lower0.5ex\hbox{$\scriptstyle<$}} \atop

%\smash{raise0.3ex\hbox{$\scriptstyle \sim$}}
%$}}
\def\lsim{\;\raise0.3ex\hbox{$<$\kern-0.75em\raise-1.1ex\hbox{$\sim$}}\;}
\def\gsim{\;\raise0.3ex\hbox{$>$\kern-0.75em\raise-1.1ex\hbox{$\sim$}}\;}
\def\no{\noindent}

\def\ce{\centerline}
\def\ve{\vfill\eject}
\def\rdots{\mathinner{\mkern1mu\raise1pt\vbox{\kern7pt\hbox{.}}\mkern2mu
 \raise4pt\hbox{.}\mkern2mu\raise7pt\hbox{.}\mkern1mu}}

\def\e e{$e^+ e^-$ }

%\input epsf

%End of Beginning Formats
%Beginning of Letter Heading

\rightline{hep-th/0206067}
\rightline{UCLA/02/TEP/12}

%\vskip1.5cm

\ce{\bf ON $q-SU(3)$ GLOBAL GAUGE THEORY}
\vskip.5cm

\ce{\it Robert J. Finkelstein}
\vskip.3cm

\ce{Department of Physics and Astronomy}
\ce{University of California, Los Angeles, CA 90095-1547}
\vskip1.0cm

\no{\bf Abstract.}  We study the replacement of $SU(3)$ by $SU_q(3)$ 
in standard gauge theories.  At the level of a global theory there is a
physically sensible $SU_q(3)$ formalism with measurable differences from
the $SU(3)$ theory.  

\vskip3.0cm

\line{{PACS Index Categories:} \hfil}  
\line{{~~81R50} \hfil}
\line{{~~81V05} \hfil}

\ve

\line{{\bf 1. Introduction.} \hfil}
\vskip.3cm

It is perhaps possible to construct a $q$-gauge theory by replacing the Lie group of
the standard gauge theory by the corresponding $q$-group.  Since the so obtained
theory has more degrees of freedom than the theory from which it is
derived, it may be possible to interpret these new degrees of freedom as
descriptive of a non-locality that is implemented by the appearance of 
solitons in the derived theory.$^1$  It also turns out that the original Lie algebra gets replaced by two dual algebras, the first
of which lies close to and approximates the original Lie algebra in the
correspondence limit $(q=1)$ while the second algebra is new and disappears
in the same limit.$^1$

Corresponding to the two dual algebras we may consider two descriptions of the
same particle.  In a macroscopic description, corresponding to the first
algebra, the physical particles are viewed as pointlike, while the same
particles in the complementary microscopic description corresponding to the
second algebra, are viewed as solitons.  We shall refer to the first and
second algebra as external and internal algebras.

In an earlier note we have discussed the Weinberg-Salam theory in terms of
the external algebra of $SU_q(2).^{~2}$  Here we are interested in the external
algebra of $SU_q(3)$ that approximates the standard $SU(3)$ flavor description
of the hadrons.  One may speculate that the internal algebra of $SU_q(3)$ provides the complementary description of the same hadrons in terms of quarks and
gluons while the internal algebra of $SU_q(2)$ describes the solitonic structure
of the massive electroweak particles.

We are here using the language of Lie groups rather than Hopf algebras since
we want to emphasize a correspondence limit with standard theory in which
``internal algebra" corresponds to the usual Lie group and ``external algebra"
corresponds to the usual Lie algebra.  Since the standard Lie group may be
obtained by integrating its Lie algebra, all degrees of freedom of the
standard theory are already exposed in the Lie algebra.  That is not true in the
$q$-theory.  Therefore in limiting this paper to the external algebra, we are
describing only certain perturbative aspects of the full theory.  The following
discussion of $SU_q(3)$ lies roughly at the same level as the ``Eightfold Way"
description of $SU(3)$.  We have not attempted to construct a local
$SU(3)$ theory.
\ve

\line{{\bf 2. The $SU(3)$ and the External $SU_q(3)$ Algebra.} \hfil}
\vskip.3cm

Let us introduce the generating matrices
$$
\eqalign{\bar E^\alpha &= \left(\matrix{0 & 1 & 0 \cr 0 & 0 & 0 \cr 0 & 0 & 0 \cr}
\right) \qquad\bar E^\beta = \left(\matrix{0 & 0 & 0 \cr 0 & 0 & 1 \cr
0 & 0 & 0 \cr} \right) \qquad\bar E^\theta = \left(\matrix{0 & 0 & 1 \cr
0 & 0 & 0 \cr 0 & 0 & 0 \cr} \right) \cr
H^\alpha &= \left(\matrix{1 & 0 & 0 \cr 0 & -1 & 0 \cr 0 & 0 & 0 \cr} \right)
\quad H^\beta = \left(\matrix{0 & 0 & 0 \cr 0 & 1 & 0 \cr
0 & 0 & -1 \cr} \right) \quad H^\theta = \left(\matrix{
1 & 0 & 0 \cr 0 & 0 & 0 \cr 0 & 0 & -1 \cr} \right) 
\cr} \eqno(2.1)
$$
\no The commutation rules of the $SU(3)$ algebra are
$$
\eqalign{
&(\bar E^\alpha, E^\alpha) = H^\alpha \cr
&(\bar E^\beta, E^\beta) = H^\beta \quad {\rm (2.2)} \cr
&(\bar E^\theta, E^\theta) = H^\theta \cr
\hfil \cr} \quad
\eqalign{
&(H^\alpha,\bar E^\alpha) = 2\bar E^\alpha \cr
&(H^\beta, \bar E^\beta) = 2\bar E^\beta \quad {\rm (2.3)} \cr
&(H^\alpha,\bar E^\beta) = -\bar E^\beta \cr
&(H^\beta,\bar E^\alpha) = -\bar E^\alpha \cr} \quad
\eqalign{
&(\bar E^\alpha,\bar E^\beta) = \bar E^\theta \cr
&(\bar E^\alpha,E^\theta) = -E^\beta  \quad {\rm (2.4)} \cr
&(\bar E^\beta,E^\theta) = E^\alpha \cr
\hfil \cr} 
$$
\no where the bar indicates Hermitian adjoint.  The remaining relations are
obtained by Hermitian conjugation.  In the Chevalley basis these may be written as
$$
\eqalignno{(\bar E^i,E^j) &= \delta^{ij}H^i \qquad i,j = 1,2 & (2.5a) \cr
(H^i,\bar E^j) &= A^{ji}\bar E^j & (2.5b) \cr
(H^i,H^j) &= 0 & (2.5c) \cr}
$$
\no plus Hermitian conjugate relations.  Here $A^{ji}$ is the Cartan matrix
$$
A^{ji} = \left(\matrix{2 & -1 \cr -1 & 2 \cr}\right) \eqno(2.5d)
$$
\no To complete the formulation (2.5) one should add the Serre and the Jacobi
relations.

We shall understand by the $SU_q(3)$ algebra the relations (2.5) with 
(2.5a) replaced by$^3$
$$
(\bar E^i,E^j) = \delta^{ji}[H^i]_q \eqno(2.6)
$$
\no plus appropriately modified Serre relations with which we need not be
concerned.  Here
$$
[H^i]_q = {q^{H^i}-q_1^{H^i}\over q-q_1} \qquad
q_1 = q^{-1} \eqno(2.7)
$$

We connect with the usual notation of particle physics by setting$^4$
$$
\eqalign{&E^1 = T^- \cr
&E^2 = U^- \cr} \qquad
\eqalign{ 
{\rm (2.8a)} \cr} \qquad \quad
\eqalign{&H^1 = 2T^3 \cr
&H^2 = {3\over 2} Y-T^3 \cr} \qquad
\eqalign{ 
{\rm (2.8b)} \cr}
$$
\no where $T,U$, and $Y$ have their usual meaning as isotopic spin, $U$-spin
and hypercharge.  Then
$$
\eqalignno{(T^+,T^-) &= 2T^3 & (2.9) \cr
(U^+,U^-) &= {3\over 2}~ Y-T^3 & (2.10) \cr}
$$
\no where
$$
\eqalign{T^- &= \bar T^+ \cr
U^- &= \bar U^+ \cr
V^- &= \bar V^+ \cr}   \eqno(2.11)
$$
\vskip.5cm

\line{{\bf 3. Selection Rules on Ladder Operators.} \hfil}
\vskip.3cm

Choose a basis in which $H^1$ and $H^2$ are diagonal with eigenvalues $m_1$
and $m_2$.  In this basis we have by (2.5)
$$
\langle m_1^\prime m_2^\prime|H^1E^1-E^1H^1 + 2E^1|m_1m_2\rangle = 0 \eqno(3.1)
$$
\no or
$$
(m_1^\prime-m_1+2)\langle m_1^\prime m_2^\prime|E^1|m_1m_2\rangle = 0
\eqno(3.2)
$$
\no Similarly
$$
\eqalignno{
&\langle m_1^\prime m_2^\prime|H^2E^1-E^1H^2 - E^1|m_1m_2\rangle = 0 
& (3.3) \cr
&(m_2^\prime-m_2-1)\langle m_1^\prime m_2^\prime|E^1|m_1m_2\rangle = 0
& (3.4) \cr}
$$
\no Therefore
$$
\eqalign{
\langle m_1^\prime m_2^\prime|E^1|m_1m_2\rangle = 0 \qquad \quad
&{\rm unless}~~m_1^\prime -m_1 = -2 \cr
&{\rm and}~~~~~m_2^\prime-m_2 = 1 \cr} \eqno(3.5)
$$
\no The corresponding selection rule on $E^2$ is
$$
\eqalign{
\langle m_1^\prime m_2^\prime|E^2|m_1m_2\rangle = 0 \qquad \quad
&{\rm unless}~~m_1^\prime-m_1 = 1 \cr
&{\rm and}~~~~~m_2^\prime-m_2 = -2 \cr} \eqno(3.6)
$$
\no The equations (3.5) and (3.6) may be combined as
$$
\eqalign{\langle\vec m-\vec\Delta^i|E^i|\vec m\rangle = 0 \qquad i=1,2
\qquad\quad &{\rm unless}~~\vec\Delta^1 = (2,-1) \cr
&{\rm or}~~~~~~~\vec\Delta^2 = (-1,2) \cr} \eqno(3.7)
$$
\vskip.5cm

\line{{\bf 4. Matrix Elements of the Ladder Operators.} \hfil}
\vskip.3cm

We have by (2.6)
$$
(\bar E^i,E^i) = [H^i]_q \eqno(4.1)
$$
\no where henceforth the subscript $q$ in $[H^i]_q$ is not indicated.
Hence by (3.7)
$$
\eqalign{\langle\vec m|\bar E^i|\vec m-\vec\Delta^i\rangle
\langle\vec m-\vec\Delta^i|E^i|\vec m\rangle -
&\langle\vec m|E^i|\vec m+\vec\Delta^i\rangle\langle\vec m+\vec\Delta^i|
\bar E^i|\vec m\rangle \cr
&=\langle\vec m|[H^i]|\vec m\rangle \cr} \eqno(4.2)
$$
\no or
$$
|\langle\vec m-\vec\Delta^i|E^i|\vec m\rangle|^2-|\langle\vec m|E^i|
\vec m+\vec\Delta^i\rangle|^2 = \langle\vec m|[H^i]|\vec m\rangle
\eqno(4.3)
$$
\no and
$$
|\langle\vec m|E^i|\vec m+\vec\Delta^i\rangle|^2-|\langle\vec m+\vec\Delta^i|
E^i|\vec m+2\vec\Delta^i\rangle|^2 =
\langle\vec m+\vec\Delta^i|[H^i]|\vec m+\vec\Delta^i\rangle \eqno(4.4)
$$
\no Define
$$
f_i(\vec m) = |\langle\vec m|E^i|\vec m+\vec\Delta^i\rangle|^2 \eqno(4.5)
$$
\no Then rewrite (4.4):
$$
f_i(\vec m)-f_i(\vec m+\vec\Delta^i) = g_i(\vec m+\vec\Delta^i) \eqno(4.6)
$$
\no where
$$
g_i(\vec m) = \langle\vec m|[H^i]|\vec m\rangle \eqno(4.7)
$$
\no Eq. (4.6) implies the following sequence
$$
\eqalign{&f_i(\vec m+\vec\Delta^i)-f_i(\vec m + 2\vec\Delta^i)
= g_i(\vec m+2\vec\Delta^i) \cr
&\qquad \qquad \qquad {\bf .}~~{\bf .}~~{\bf .} \cr
&f_i(\vec m+(s-1)\vec\Delta^i)-f_i(\vec m+s\vec\Delta^i) 
= g_i(\vec m+s\vec\Delta^i) \cr} \eqno(4.8)
$$
\no and adding the equations in the sequence (4.6)-(4.8) one finds
$$
f_i(\vec m) -f_i(\vec m+s\vec\Delta^i) =
\sum^s_{t=1} g_i(\vec m+t\vec\Delta^i) \eqno(4.9)
$$

To evaluate the right side of (4.9) note
$$
\eqalignno{\langle\vec m + t\vec\Delta^i|H^i|\vec m + t\vec\Delta^i\rangle &=
m_i+t\Delta^i_i & (4.10a) \cr
&= m_i+2t & (4.10b) \cr}
$$
\no by (3.7).  Then by (2.7)
$$
g_i(\vec m + t\vec\Delta^i) =
{q^{m_i+2t}-q_1^{m_i+2t}\over q-q_1} \eqno(4.11)
$$
\no Note also
$$
\eqalign{
\sum^s_{t=1} q^{m_i+2t} &= q^{m_i+2}\sum^{s-1}_{t=0}(q^2)^t \cr
&= q^{m_i+2}\langle s\rangle_{q^2} \cr} \eqno(4.12)
$$
\no where $\langle s\rangle_{q^2}$ is defined by (4.12) or by
$$
\eqalign{\langle s\rangle_q &= {q^s-1\over q-1} \cr
&= 1+q+ \ldots +q^{s-1} \cr} \eqno(4.13)
$$
\no Here $\langle s\rangle_q$ is the basic number corresponding to
$s$.  Then by (4.11)
$$
\sum^s_{t=1} g_i(\vec m+t\vec\Delta_i) =
[q^{m_i+2}\langle s\rangle_{q^2}-q_1^{m_i+2}\langle s\rangle_{q_1^2}]
(q-q_1)^{-1} \eqno(4.14)
$$
\no Since
$$
\langle s\rangle_{q_1} = q_1^{s-1}\langle s\rangle_q \eqno(4.15)
$$
\no we have
$$
\eqalignno{\sum^s_{t=1} g_i(\vec m+t\vec\Delta_i) &= (q^{m_i+2}-q_1^{m_i+2s})
\langle s\rangle_{q^2}(q-q_1)^{-1} & (4.16) \cr
&= q_1^{m_i+2s}\biggl[{(q^2)^{m_i+s+1}-1\over q^2-1}\biggr]
{q^2-1\over q-q_1} \langle s\rangle_{q^2} & (4.17) \cr}
$$
\no Then by (4.9) and the preceding equation
$$
f_i(\vec m)-f_i(\vec m + s\vec\Delta_i) = q_1^{m_i+2s-1}
\langle m_i+s+1\rangle_{q^2}\langle s\rangle_{q^2} \eqno(4.18)
$$
\no The function $f_i(m+s\Delta)$ given by (4.18) satisfies the difference
Eq. (4.6) for $s=1$.  Although (4.18) was obtained for $s$ integral, it has
meaning for non-integral values of $s$.  Set $\vec m=\vec 0$.  Then
$$
f_i(s\vec\Delta_i) = f_i(\vec 0)-q_1^{2s-1}\langle s+1\rangle_{q^2}
\langle s\rangle_{q^2} \eqno(4.19)
$$

The second term of (4.19) is negative and grows in absolute value
as $s$ increases.  Since both
the left side of (4.19) and $f_i(0)$ are positive, the maximum value of $s$, say
$j_i$, is given by
$$
f_i(0) = q_1^{2j_i-1}\langle j_i\rangle_{q^2}\langle j_i+1\rangle_{q^2}
\eqno(4.20)
$$
\no where $j$ may not be integral.  By (4.19) and (4.20)
$$
f_i(s\vec\Delta_i) = q_1^{2j_i-1}\langle j_i\rangle_{q^2}\langle j_i+1\rangle_{q^2}-
q_1^{2s-1}\langle s\rangle_{q^2}\langle s+1\rangle_{q^2}
\eqno(4.21)
$$
\no and
$$
f_i(j_i\vec\Delta_i) = 0 \eqno(4.22)
$$
\no Also, if $s=-j_i-1$, we have
$$
f_i[(-j_i-1)\vec\Delta_i] = q_1^{2j-1}\langle j_i\rangle_{q^2}\langle j_i+1\rangle_{q^2}
-q_1^{-2j-3}\langle -j_i-1\rangle_{q^2}\langle -j_i\rangle_{q^2} \eqno(4.23)
$$
\no and the second term of the preceding equation gives
$$
q_1^{-2j-3}\langle -j_i-1\rangle_{q^2}\langle -j_i\rangle_{q^2} =
q_1^{-2j-3}(q^2)^{-j-1}(q^2)^{-j}\langle j_i+1\rangle_{q^2} \langle j_i\rangle_{q^2}
\eqno(4.24)
$$
\no since
$$
\langle -n\rangle_q = -q^{-n}\langle n\rangle_q \eqno(4.25)
$$
\no By (4.24) and (4.25)
$$
f_i[(-j_i-1)\vec\Delta_i] = 0 \eqno(4.26)
$$
\no But
$$
f_i(s\vec\Delta_i) = |\langle s\vec\Delta_i|E_i|(s+1)\vec\Delta_i\rangle|^2 \eqno(4.27)
$$
\no by (4.5).  Then
$$
|\langle j_i\vec\Delta_i|E_i|(j_i+1)\vec\Delta_i\rangle|^2 = 
|\langle(j_i+1)\vec\Delta_i|\bar E_i|j_i\vec\Delta_i\rangle|^2 = 0 \eqno(4.28)
$$
\no and
$$
|\langle(-j_i-1)\vec\Delta_i|E_i|(-j_i)\vec\Delta_i\rangle|^2 =
|\langle(-j_i)\vec\Delta_i|\bar E_i|(-j_i-1)\vec\Delta_i\rangle|^2 = 0
\eqno(4.29)
$$
\no By (4.28) the state $(j_i+1)$ cannot be reached by a raising operator,
and by (4.29) the state $-j_i-1$ is inaccessible with a lowering operator.

The number of steps $(2j)$ between $-j$ and $+j$ is either an odd or even integer and $j$ is either half integer or integer.  By (4.21) and (4.27) we have
$$
|\langle s\vec\Delta_i|E_i|(s+1)\vec\Delta_i\rangle|^2 =
q_1^{2j_i-1}\langle j_i\rangle_{q^2}\langle j_i+1\rangle_{q^2}-q_1^{2s-1}
\langle s\rangle_{q^2}\langle s+1\rangle_{q^2} \eqno(4.30)
$$
\no In the $q=1$ limit we have the familiar
$$
|\langle s\vec\Delta_i|E_i|(s+1)\vec\Delta_i\rangle|^2 = 
(j-s)(j+s+1) \eqno(4.31)
$$
\ve

\line{{\bf 5. The Irreducible Representations of the External Algebra.} \hfil}
\vskip.3cm

Since the rank of the algebra is two, the irreducible representations may
be labeled by integers $(n_1,n_2) = 2(j_1,j_2)$.  In terms of $n_1$ and
$n_2$ the dimensionality of the irreducible representation $(n_1,n_2)$ is
$$
D = {1\over 2} (n_1+1)(n_2+1)(n_1+n_2+2) \eqno(5.1)
$$

Instead of labeling the states by the eigenvalues of $H_1$ and $H_2$ it is
customary in discussions of flavor to use $Y$, the hypercharge, and $T_3$, one
component of the isotopic spin, as shown in (2.8b).  The states of an
irreducible representation may be labeled by the eigenvalues, $y$ and
$t_3$, of $Y$ and $T_3$ and plotted in the $(y,t_3)$ plane.  The perimeter
of the two-dimensional set of points belonging to the irreducible representation $(n_1,n_2)$ is composed of segments whose lengths are determined by the
number of successive steps connecting extreme accessible states.  Thus the
length of a boundary segment is either $n_1$ or $n_2$ steps.$^4$  The boundary
is triangular if $n_1$ or $n_2$ vanishes.  Otherwise it will be hexagonal.

One also has by (2.4)
$$
(E^\alpha,E^\beta) =  -E^\theta \eqno(5.2)
$$
\no or
$$
\eqalign{
\langle\vec m|E^\alpha|&\vec m+\vec\Delta^\alpha\rangle\langle\vec m\vec +\Delta^\alpha|E^\beta|\vec m +\vec\Delta^\alpha +\vec\Delta^\beta\rangle \cr 
&-\langle\vec m|E^\beta|\vec m+\vec\Delta^\beta\rangle\langle\vec m+\vec\Delta^\beta|E^\alpha|\vec m+\vec\Delta^\beta + \vec\Delta^\alpha\rangle
= -\langle\vec m|E^\theta|\vec m+\vec\Delta^\alpha + \vec\Delta^\beta\rangle \cr} \eqno(5.3)
$$
\no Hence $E^\theta$ displaces in the direction
$$
\vec\Delta^\theta = \vec\Delta^\alpha + \vec\Delta^\beta \eqno(5.4)
$$
\no If $\vec\Delta^\theta$ is a boundary displacement, the number of steps
is again either $j_1$ or $j_2$.

By (2.8b) the eigenvalues $m^1$ and $m^2$ of $H^1$ and $H^2$ are related
to the corresponding eigenvalues of $T^3$ and $Y$ by
$$
\eqalign{
m^1 &= 2t^3 \cr
m^2 &= {3\over 2} y-t^3 \cr} \eqno(5.5)
$$
\no Then for the displacement $\vec\Delta^1 = (2,-1)$ one has
$$
\left(\matrix{2 \cr -1 \cr} \right) =
\left(\matrix{ \Delta m_1 \cr \Delta m_2 \cr}\right) =
\left(\matrix{2\Delta t^3 \cr {3\over 2}\Delta y-\Delta t^3 \cr} \right)
\eqno(5.6)
$$
\no or
$$
\Delta y = 0 \eqno(5.7)
$$
\no Thus $\vec\Delta^1$ lies in the direction $\Delta y = 0$.

For the displacement $\vec\Delta^2 = (-1,2)$ we have
$$
\left(\matrix{ -1 \cr 2 \cr} \right) =
\left(\matrix{ 2\Delta t^3 \cr {3\over 2}\Delta y - \Delta t^3 \cr}\right)
\eqno(5.8)
$$
\no Hence
$$
\Delta y = -2\Delta t^3
$$
\no and $\vec\Delta^2$ has slope -2 in the $ty$-plane.
\vskip.5cm

\line{{\bf 6. Matrix Elements and their Relation to $SU_q(2)$.} \hfil}
\vskip.3cm

The values of the matrix elements according to (4.30) lie close to the
corresponding values for $SU(3)$ if $q$ is near unity.  Still there would be
systematic deviations from $SU(3)$ predictions resulting from the
slightly different matrix elements.  

Our earlier discussion of $q$-electroweak was based on the commutator:
$$
\eqalign{(J_+,J_-) &= q_1^{2j-1}[2J_3] \cr
(J_3,J_+) &= J_+ \qquad (J_3,J_-) = -J_- \cr} \eqno(6.1)
$$
\no instead of
$$
(J_+,J_-) = [J_3] \eqno(6.2)
$$
\no corresponding to (2.6) used here.

Eq. (6.1) may be put in the form (6.2) by the following rescaling
$$
\eqalign{J_\pm &= \lambda\hat J_\pm \cr
J_3 &= \hat J_3 \cr
q &= \hat q^{1/2} \cr} \eqno(6.3)
$$
\no where
$$
\lambda^2 = \langle 2\rangle_{\hat q} \hat q_1^j
$$
\no Then
$$
(\hat J_+,\hat J_-) = [\hat J_3]_{\hat q} \eqno(6.4)
$$
\no the form used in the present paper.

The general result of the earlier discussion of $SU_q(2)$ is that the
fundamental matrix representations of $SU(2)$ and $SU_q(2)$ agree and that
the differences first appear in the adjoint representation.  We now examine
this question for $SU(3)$ and $SU_q(3)$.

Let us consider the triplet and octet representations for which
$(n_1,n_2)$ are $(0,1)$, $(1,0)$, and $(1,1)$.

\vskip.5cm

\line{{\bf 7. The Fundamental and Adjoint Representations.} \hfil}
\vskip.3cm

The fundamental representations of $SU(3)$ may be labelled $(1,0)$ and $(0,1)$
corresponding to $(j_1,j_2) = (1/2,0)$ and $(0,1/2)$.  These are three-dimensional spaces spanned by 3 vectors that are denoted in the quark notation by $u,d$, and $s$ with $(t,y)$ coordinates $(1/2,1/3)$, $(-1/2,1/3)$,
and $(0,-2/3)$ in the triplet $(1,0)$.

In $(m_1,m_2)$ coordinates we have 
$$
\eqalign{u &= (1,0) \cr
d &= (-1,1) \cr
s &= (0,-1) \cr} \qquad
\hbox{and} \qquad
\eqalign{\Delta_1 &= (2,-1) \cr
\Delta_2 &= (-1,2) \cr
\hfil \cr}
\eqno(7.1)
$$
\no Here the symbols $u,d$, and $s$ are to be regarded as two-dimensional
vectors and
$$
\eqalignno{u &= d+\Delta_1 & (7.2) \cr
d &= s+\Delta_2 & (7.3) \cr}
$$
\no The non-vanishing elements of $E_1$ and $E_2$ are
$$
\eqalignno{\langle d|E_1|u\rangle &= \langle d|E_1|d+\Delta_1\rangle &
(7.4) \cr
\langle s|E_2|d\rangle &= \langle s|E_2|s+\Delta_2\rangle & (7.5) \cr}
$$
\no Then
$$
\langle s|E_3|u\rangle = \langle s|E_2E_1|u\rangle =
\langle s|E_3|s+\Delta_1+\Delta_2\rangle
$$
\no where
$$
E_3 = (E_2,E_1)
$$
\no These connections are shown in the familiar Fig. 1:

\ve 

%%%%%%%%%%%%%%%%%%%%%%%%%%%%%%
%FIGURE
\vskip .3 cm 
\centerline{\epsfxsize 1.2 truein \epsfbox{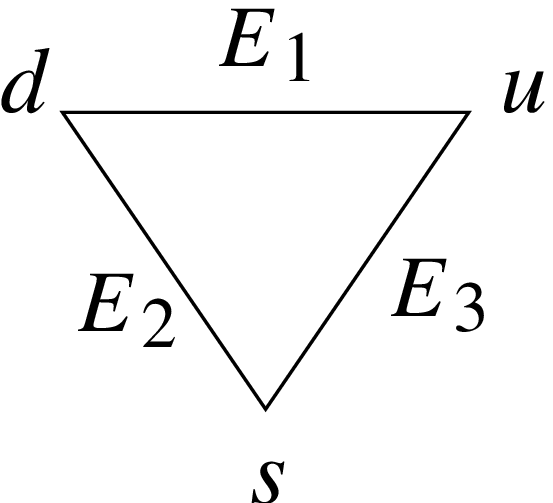}}
{{\bf Figure 1:} The Quark Representation}
%%%%%%%%%%%%%%%%%%%%%%%%%
\vskip .8 cm 

The matrix elements $\langle d|E_1|u\rangle$ and $\langle s|E_2|d\rangle$
are given by (4.18) and (4.5) in the following form:
$$
\eqalign{&|\langle \vec m+s\vec\Delta_i|E_i|\vec m + (s+1)
\vec\Delta_i\rangle|^2 \cr
&= q_1^{m_i}\bigl[q_1^{2j_i-1}\langle m_i+j_i+1\rangle_{q^2}
\langle j_i\rangle_{q^2}-q_1^{2s-1}\langle m_i+s+1\rangle_{q^2}
\langle s\rangle_{q^2}\bigr] \cr}
\eqno(7.7)
$$
\no In both cases $E_i$ connects boundary states in the triplet multiplet.
Then $2j=1$ in both cases.  The (lowest, highest) values of $s$ are
$(-j,j) = (-1/2,1/2)$.  Take $s=-1/2$.  Then (7.7) reads
$$
|\langle\vec m-{1\over 2}\vec\Delta_i|E_i|\vec m + {1\over 2}\vec\Delta_i\rangle|^2 = q_1^{m_i}\biggl[\langle m_i + {3\over 2}\rangle_{q^2}
\langle{1\over 2}\rangle_{q^2}-q_1^{-2}
\langle m_i+{1\over 2}\rangle_{q^2}\langle-{1\over 2}\rangle_{q^2}\biggr]
\eqno(7.8)
$$
\no In the cases of interest we have
$$
\langle d|E_1|u\rangle = \langle\vec m-{1\over 2}\vec\Delta_1|E_1|
\vec m + {1\over 2}\vec\Delta_1\rangle \eqno(7.9)
$$
\no and
$$
\langle s|E_2|d\rangle = \langle\vec m-{1\over 2}\vec\Delta_2|E_2|\vec m +
{1\over 2}\vec\Delta_2\rangle \eqno(7.10)
$$
\no By (7.9)
$$
\vec m - {1\over 2}\vec\Delta_1 = d \eqno(7.11)
$$
\no or
$(m_1m_2) - {1\over 2}(2,-1) = (-1,1)$ and
$$
m_1=0 \eqno(7.12)
$$
\no Then by (7.8) and (7.12)
$$
|\langle d|E_1|u\rangle|^2 = \bigl\langle{3\over 2}\bigr\rangle_{q^2}
\bigl\langle{1\over 2}\bigr\rangle_{q^2}-q^2\bigl\langle{1\over 2}\bigr\rangle_{q^2}\bigl\langle-{1\over 2}\bigr\rangle_{q^2} \eqno(7.13)
$$

Similarly by (7.10)
$$
\vec m - {1\over 2}\vec\Delta_2 = s \eqno(7.14)
$$
\no and
$$
m_2 = 0 \eqno(7.15)
$$
\no and
$$
|\langle s|E_2|d\rangle|^2 = |\langle d|E_1|u\rangle|^2 \eqno(7.16)
$$
\no just as found in (7.13).  This common expression reduces as follows:
$$
\eqalignno{|\langle d|E_1|u\rangle|^2 &=
|\langle s|E_2|d\rangle|^2 = \bigl\langle{1\over 2}\bigr\rangle_{q^2}
\biggl[\bigl\langle{3\over 2}\bigr\rangle_{q^2} + q\bigl\langle{1\over 2}\bigr\rangle_{q^2}\biggr] & (7.17) \cr
&= \biggl({1\over\langle 2\rangle_q}\biggr)^2 \bigl[\langle 3\rangle_q +
q\langle 1\rangle_q\bigr] & (7.18) \cr}
$$
\no by (4.25) and
$$
\bigl\langle{n\over 2}\bigr\rangle_{q^2} =
{1\over\langle 2\rangle_q} \langle n\rangle_q \eqno(7.19)
$$
\no But
$$
\langle 3\rangle_q + q\langle 1\rangle_q =
(1+q+q^2) + q = (1+q)^2 = \langle 2\rangle_q^2 \eqno(7.20)
$$
\no Hence
$$
|\langle d|E_1|u\rangle|^2 = |\langle s|E_2|d\rangle|^2 =
|\langle s|E_3|u\rangle|^2 = 1 \eqno(7.21)
$$
\no Thus all of these matrices in the fundamental representation are unchanged
in going from $SU(3)$ to $SU_q(3)$.
\vskip.5cm

\line{{\bf 8. The Adjoint Representation (1,1).} \hfil}
\vskip.3cm

There are now eight states.  We display the $SU(3)$ baryon octet in the
$ty$-plane on the left and the $(m_1,m_2)$ coordinates below.
\vskip.5cm

%%%%%%%%%%%%%%%%%%%%%%%%%%%
%FIGURE
\vskip .3 cm 
\centerline{\epsfxsize 2.5 truein \epsfbox{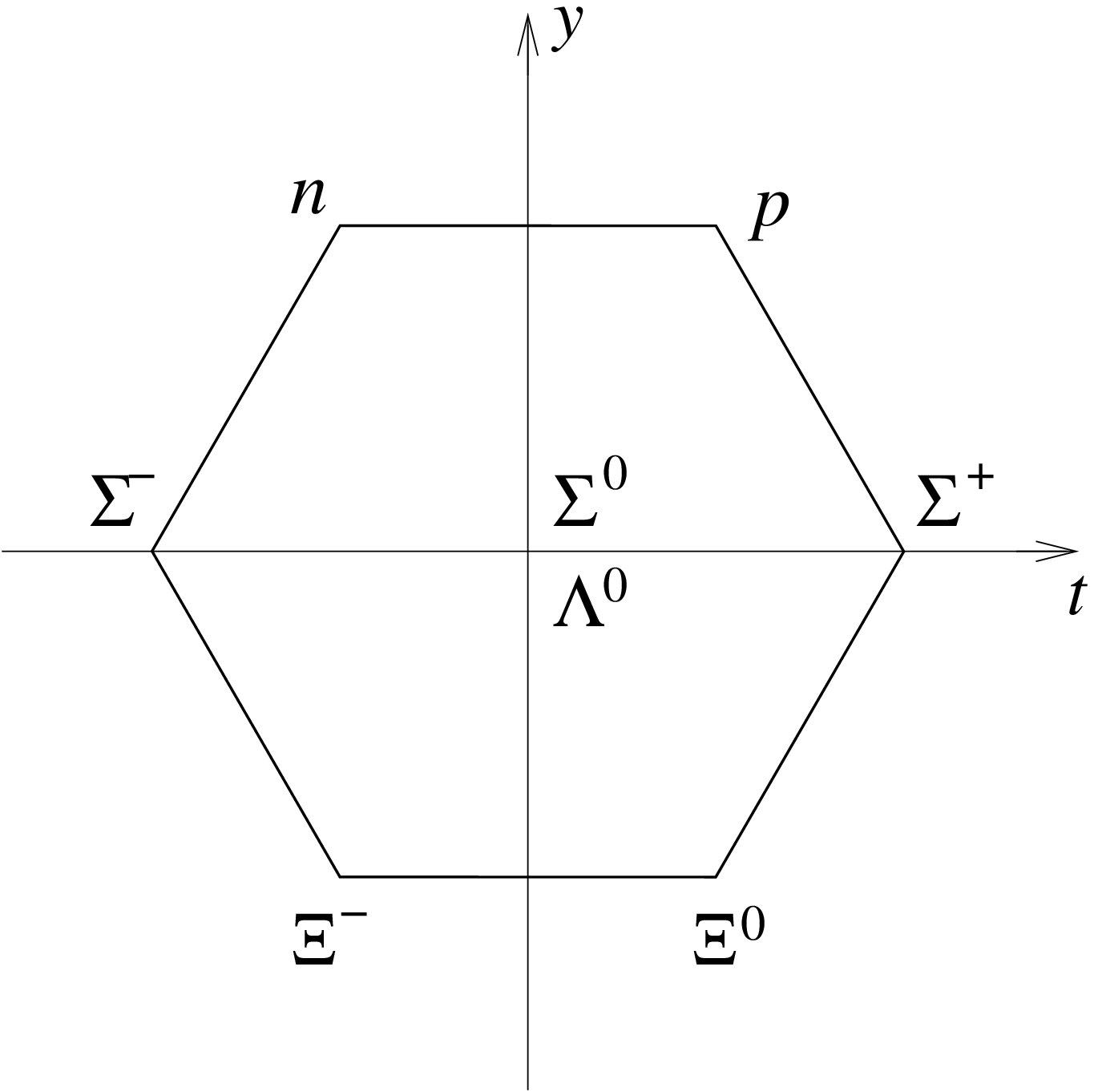}}
%{\bf Figure 2:} The Octet Representation and the $(m_1m_2)$ Coordinator
%%%%%%%%%%%%%%%%%%%%%%%%%
\vskip .8 cm 

% height2pt&\omit&\omit\cr
$$
\vbox{\offinterlineskip
\halign{# & \vrule \strut ~ # && \hfil # \hfil \cr
\hfil & $n$ & $p$ & $\Sigma^-$ & $\Sigma^+$ & $\Xi^-$ & $\Xi^o$ & $\Sigma
^o$ & $\Lambda^o$ \cr
\noalign{\hrule}
$m_1$ & $-1$ & $1$ & $-2$ & $2$ & $-1$ &  $1$ & $0$ & $0$ \cr
$m_2$ & $2$ & $1$ & $1$ & $-1$ & $-1$ & $-2$ & $0$ & $0$ \cr}}
\eqno(8.1)
$$
\vskip.5cm

{\bf Figure 2:} The Octet Representation and the $(m_1m_2)$ Coordinates

%$$
%\vbox{\halign{<preamble> \cr
%\hfil & n &p &\Sigma^- &\Sigma^+ &\Xi^- &\Xi^o &\Sigma^o &\Lambda^o \cr
%m_1 &-1 &1 &-2 &2 &-1 &1 &0 &0 \cr
%m_2 &2 &1 &1 &-1 &-1 &-2 &0 &0 \cr}}
%\eqno(8.1)
%$$
 
\vskip.5cm

%\vskip2.0in

\no The non-vanishing matrix elements include
$$
\eqalign{\langle n|E_1|p\rangle &= \langle n|E_1|n+\Delta_1\rangle \cr
\langle\Sigma^+|E_2|p\rangle &= \langle\Sigma^+|E_2|\Sigma^++\Delta_2\rangle \cr
\langle\Xi^o|E_3|\Sigma^+\rangle &= \langle\Xi^o|E_3|\Xi^o + \Delta_3\rangle \cr
\langle\Xi^-|E_1|\Xi^o\rangle &= \langle\Xi^-|E_1|\Xi^o + \Delta_1\rangle \cr
\langle\Xi^-|E_2|\Sigma^o\rangle &= \langle\Xi^-|E_2|\Xi^- + \Delta_2\rangle \cr
\langle\Sigma^-|E_3|n\rangle &= \langle\Sigma^-|E_3|\Sigma^- + \Delta_3\rangle \cr} \eqno(8.2)
$$
\no where $\Delta_1$ and $\Delta_2$ are given in (3.7).  In addition we have
matrix elements connecting boundary states with $\Lambda^o$ and $\Sigma^o$.
For example
$$
\eqalign{\langle\Lambda^o|E^1|\Sigma^+\rangle &=
\langle\Lambda^o|E^1|\Lambda^o + \Delta_1\rangle \cr
\langle\Xi^o|E^2|\Lambda^o\rangle &= \langle\Xi^o|E^2|\Xi^o + \Delta_2\rangle
\cr} \eqno(8.3)
$$
\no Instead of $\Lambda^o$ we could write $\Sigma^o$.  In addition
$$
\eqalign{\langle\Xi^o|E^3|\Sigma^+\rangle &=
\langle\Xi^o|E^2E^1|\Sigma^+\rangle \cr
&= \langle\Xi^o|E^2|\Lambda^o\rangle\langle\Lambda^o|E^1|\Sigma^+\rangle \cr}
\eqno(8.4)
$$
\no These matrix elements are also determined by (7.7).

If $j_i = 1/2,~s = -1/2$, one has (7.8) again.

To evaluate one needs $m_i$ in each case, i.e. the $m_i$ component of the
left-hand state.  For example
$$
\vec m + s\vec\Delta_1 = n
$$
\no or
$$
m_1 = 0 \eqno(8.5)
$$
\no and
$$
\vec m + s\vec\Delta_2 = \Sigma^+
$$
\no or
$$
m_2 = 0 \eqno(8.6)
$$
\no In both cases one gets the earlier result found in (7.16), namely
$$
\eqalign{|\langle n|E_1|p\rangle|^2 &= |\langle\Sigma^+|E_2|p\rangle|^2 \cr
&= \biggl\langle{3\over 2}\biggr\rangle_{q^2}
\biggl\langle{1\over 2}\biggr\rangle_{q^2}-q^2\biggl\langle{1\over 2}\biggr\rangle_{q^2}\biggl\langle-{1\over 2}\biggr\rangle_{q^2} = 1 \cr}
\eqno(8.7)
$$

\no Again, matrix elements connecting boundary states are unchanged from their
$SU(3)$ values.

There are in addition to the doublet boundary states, triplets of states
such as $\langle\Sigma^-,\Sigma^o,\Sigma^+\rangle$ for which $j=1$ and the values of
$s$ are -1,0,1, repsectively.  From these triplets one obtains matrix elements
connecting the boundary states with $\Lambda^o$ and $\Sigma^o$ as follows.

Set $\vec m = \Sigma^o$ in (7.7).  Then
$$
|\langle s\Delta_i|E_i|(s+1)\Delta_i\rangle|^2 = 
q_1\langle 2\rangle_{q^2}-q_1^{2s-1}\langle s+1\rangle_{q^2}
\langle s\rangle_{q^2} \eqno(8.8)
$$
\no This matrix element vanishes for $s=1$.  For the other possibilities
($s=0$ and $s=-1$) the second term vanishes.  Hence
$$
\eqalignno{|\langle\Sigma^o|E_1|\Sigma^+\rangle|^2 &= q+q_1 \qquad
\hbox{if} \qquad s=0 & (8.9) \cr
\noalign{\hbox{and}}
|\langle\Sigma^-|E_1|\Sigma^o\rangle|^2 &= q+q_1 \qquad \hbox{if} \qquad
s=-1 & (8.10) \cr}
$$

One sees that only the matrix elements connecting boundary states with
$\Lambda^o$ and $\Sigma^o$ exhibit a difference between $SU(3)$ and $SU_q(3)$.

The corresponding relations for $E_2$ are
$$
\eqalignno{|\langle\Sigma^o|E_2|n\rangle|^2 &= (q+q_1) \qquad \hbox{if}
\qquad s=0 & (8.11) \cr
|\langle\Xi^o|E_2|\Sigma^o\rangle|^2 &= (q+q_1) \qquad \hbox{if}
\qquad s=-1 & (8.12) \cr}
$$
\no Finally
$$
\eqalign{|\langle\Sigma^o|E_2E_1|p\rangle|^2 &=
|\langle\Sigma^o|E_2|n\rangle\langle n|E_1|p\rangle|^2 \cr
&= q+q_1 \cr}
\eqno(8.13)
$$
\no We have taken over the $SU(3)$ assignments of $H$ to label the octet states
of $SU(3)_q$.  We may check these assignments by use of (4.1).  For example
$$
|\langle p-\Delta_1|E_1|p\rangle|^2 - |\langle p|E_1|p+\Delta_1\rangle|^2 =
\langle p|[H_1]|p\rangle \eqno(8.14)
$$
\no where
$$
|\langle p-\Delta_1|E_1|p\rangle|^2 = |\langle n|E_1|p\rangle|^2 = 1 \eqno(8.15)
$$
\no by (8.7) and
$$
|\langle p|E_1|p+\Delta_1\rangle|^2 = 0 \eqno(8.16)
$$
\no Then by (8.14)
$$
\langle p|[H_1]|p\rangle = [H_1(p)] = 1 \eqno(8.17)
$$
\no or
$$
H_1(p) = 1 \eqno(8.18)
$$
\no Similarly
$$
|\langle n-\Delta_1|E_1|n\rangle|^2-|\langle n|E_1|n+\Delta_1\rangle|^2 =
\langle n|[H_1]|n\rangle \eqno(8.19)
$$
\no where
$$
|\langle n-\Delta_1|E_1|n\rangle|^2 = 0 \eqno(8.20)
$$
\no and
$$
|\langle n|E_1|n+\Delta_1\rangle|^2 = |\langle n|E_1|p\rangle|^2 = 1
\eqno(8.21)
$$
\no Then by (8.19)
$$
\langle n|[H_1]|n\rangle = [H_1(n)] = -1 \eqno(8.22)
$$
\no or
$$
H_1(n) = -1 \eqno(8.23)
$$
\no The corresponding equations for $H_2$ are, for example
$$
\eqalignno{|\langle p-\Delta_2|E_2|p\rangle|^2-|\langle p|E_2|p+\Delta_2\rangle|^2 &= \langle p|[H_2]|p\rangle & (8.24) \cr
|\langle\Sigma^+|E_2|p\rangle|^2-|\langle p|E_2|p+\Delta_2\rangle|^2 &=
\langle p|[H_2]|p\rangle & (8.25) \cr
\langle p|[H_2]|p\rangle = [H_2(p)] &= 1 \cr}
$$
\no or
$$
H_2(p) = 1 \eqno(8.26)
$$
\no Similarly
$$
\eqalignno{\langle n|[H_2]|n\rangle &= |\langle n-\Delta_2|E_2|n\rangle|^2
-|\langle n|E_2|n+\Delta_2\rangle|^2 & (8.27) \cr
&= |\Sigma^o|E_2|n\rangle|^2-0 \cr
&= q+q_1 & (8.28) \cr}
$$
\no by (8.11).  Then
$$
[H_2(n)] = q+q_1  \eqno(8.29)
$$
\no or
$$
{q^{H_2(n)}-q_1^{H_2(n)}\over q-q_1} = q+q_1 \eqno(8.30)
$$
\no and
$$
H_2(n) = 2 \eqno(8.31)
$$
\no in agreement with (8.1).  The eigenvalues of $H_i$ for the central states
may also be checked by (4.3).  For example
$$
|\langle\Sigma^o-\Delta_1|E_1|\Sigma^o\rangle|^2-|\langle\Sigma^o|E_1|\Sigma^o
+\Delta_1\rangle|^2 = \langle\Sigma^o|[H_2]|\Sigma^o\rangle \eqno(8.32)
$$
\no where
$$
\eqalign{|\langle\Sigma^o-\Delta_1|E_1|\Sigma^o\rangle|^2 &=
|\langle\Sigma^-|E_1|\Sigma^o\rangle|^2 = q+q_1 \cr
|\langle\Sigma^o|E_1|\Sigma^o+\Delta_1\rangle|^2 &=
|\langle\Sigma^o|E_1|\Sigma^+\rangle|^2 = q+q_1 \cr}
$$
\no Hence
$$
[H_1(\Sigma^o)] = 0
$$
\no and
$$
H_1(\Sigma^o) = 0 \eqno(8.33)
$$

Similarly
$$
|\langle\Sigma^o-\Delta_2|E_2|\Sigma^o\rangle|^2 -
|\langle\Sigma^o|E_2|\Sigma^o+\Delta_2\rangle|^2 =
\langle\Sigma^o|[H_2]|\Sigma^o\rangle \eqno(8.34)
$$
\no and
$$
H_2(\Sigma^o) = 0 \eqno(8.35)
$$
\vskip.5cm

\line{{\bf 9. The Physical Interpretation of $SU_q(3)$.} \hfil}
\vskip.3cm

It has been pointed out in an earlier paper$^2$ that there is a reasonable
interpretation of the formalism obtained by replacing $SU(2)$ by $SU_q(2)$
in the electroweak theory.  In discussing the same question for $SU_q(3)$
one again finds that a distinction must be made between the fundamental and
adjoint representations, based on the fact that matrices in the fundamental
representations are the same for $SU_q(3)$ and $SU(3)$ while this is not
true for the adjoint representation.

The fundamental representation of $SU_q(3)$ is 3-dimensional, and since the
matrices are the same as for $SU(3)$ one may again introduce the Gell-Mann
matrices $\lambda_i$, the antisymmetric and symmetric coefficients $f_{ijk}$
and $d_{ijk}$ respectively, and the associated $F$ and $D$ operators as
follows:$^4$
$$
\eqalign{\biggl({1\over 2}\lambda_i,{1\over 2}\lambda_j\biggr) &=
i~f_{ijk}\biggl({1\over 2}\lambda_k\biggr) \cr
\{\lambda_i,\lambda_j\} &= {2\over 3}\delta_{ij}1+2d_{ijk}\lambda_k \cr}
\eqno(9.1)
$$
\no and
$$
\eqalign{F_i &= {1\over 2}\lambda_i \cr
D_i &= {2\over 3}\Sigma d_{ijk}F_jF_k \cr}
$$
\no In terms of these one may construct the Gell-Mann--Okubo mass operator.
Since this formula depends ultimately on the $\lambda_i$ and the structure
constants that appear in the fundamental representation, and since structure
constants do not have meaning for higher dimensional representations of
$SU_q(3)$, it follows that the Gell-Mann--Okubo operator in the $q$-theory is
determined entirely by the $q$-quark representation.  That it may still be
successfully applied to other representations perhaps gives additional weight
to the quark picture.

Passing on now to the higher representations one finds that the matrix
elements depend on $q$.  These modifications of the matrix elements should
in principle be measurable.  In addition there is no obstacle to the 
construction of a global theory.

On the other hand the possibility of a local (Yang-Mills) theory depends on
the existence of structure constants in the adjoint representation.  Here is
the important distinction between the $SU_q(2)$ and the $SU_q(3)$ theories:
there are structure constants in the adjoint representation of the former
but not of the latter, and there is, therefore, a local gauge theory only in the
$SU_q(2)$ theory.

The difference between these two cases may be described by reference to Eqs. (2.5) and (2.6).  In $SU_q(2)$ but not in $SU_q(3)$, Eq. (2.6) may be replaced by (2.5a).  The argument for this difference runs as follows:

In both cases we have
$$
[H] = {q^H-q_1^H\over q-q_1} \eqno(9.5)
$$
\no Set
$$
q = e^\lambda \eqno(9.6)
$$
\no then
$$
\eqalign{[H] &= {2\sinh\lambda H\over q-q_1} \cr
&= {2\over q-q_1}\biggl[\lambda H + {(\lambda H)^3\over 3!} + \ldots \biggr]
\cr} \eqno(9.7)
$$
\no In the case of $SU_q(2)$ the adjoint representation is 3-dimensional and
the eigenvalues of $H$ are $(0,\pm,1)$.  Hence
$$
H^3 = H \eqno(9.8)
$$
\no and every odd power of $H$ reduces to $H$ since
$$
H^{2p+1} = H^{3p-p+1} = H^{p-p+1} = H \eqno(9.9)
$$
\no Then
$$
[H] = H \eqno(9.10)
$$
\no In the case of $SU_q(3)$ there are two $H_i$, the adjoint representation
is 8-dimensional, and the eigenvalues of $H_i$ are (0,0,-1,1,-1,1,-2,2) so
that $H_i^2(H_i^2-1)^2(H_i^2-4) = 0$.  The right-hand side of (9.7) is
therefore no longer linear in $H_i$ and consequently there are no structure
constants for the octet representation.

From the preceding considerations it follows that there is a viable global
theory based on $SU_q(3)$ but no corresponding local theory along the lines
of the standard Yang-Mills theory.  Within the $q$-theory envisioned in this
paper one might conjecture that the description of gluons is carried by a
local theory based on the dual (internal) algebra.  (The idea of exploiting the
dual algebra has been discussed by A. Gavrilek {\it et al.}$^5$)  Our paper,
however, has focussed on the possibly observable differences between
global $SU_q(3)$ and global $SU(3)$ and does not attempt to construct a 
local field theory.

\vskip.5cm

\line{{\bf 10. Remarks.} \hfil}
\vskip.3cm

We have had in mind the replacement of a standard point particle field theory
by a soliton field theory described in the two complementary ways defined
by the dual algebras referred to in the introduction.  In the first
(macroscopic) picture the particles are regarded as pointlike and subject
to the exterior algebra while the solitons in the complementary (microscopic)
picture are regarded as composed of preons subject to the internal (dual)
algebra.
%\vskip.3cm

This idea may be partially implemented for $SU_q(2)$ where the external
algebra leads to a modified Weinberg-Salam theory and where the corresponding
soliton theory may also be formulated.$^2$  The modified Weinberg-Salam
theory exists because there is a local version of the external algebra and a
corresponding local gauge theory along the lines of a standard Yang-Mills
theory.  This $SU_q(2)$ is gauge invariant and differs only slightly from the
standard Weinberg-Salam theory.  As shown in this paper, however, there is no
local version of external $SU_q(3)$ that permits the construction of a local
gauge theory along the lines of a standard Yang-Mills theory.
%\vskip.3cm

On the other hand one may compute the differences between global $SU(3)$ and
$SU_q(3)$ as we have done here.  These differences are necessarily small
since $q$ must be near unity and any experimental test would have to
distinguish between these differences and those associated with radiative
corrections.  The calculation of radiative corrections, however, depends on
the existence of a local theory that is also renormalizable.  Unfortunately
such a theory is not at present available.

In considering whether the absence of a renormalizable external $q$-theory can be overcome, the following two points may be worth noting:

\item{(a)} Although the local $SU_q(2)$ theory differs only slightly from
the Weinberg-Salam theory and is also gauge-invariant, it is likely that it is
not tree-unitary.$^6$  As with other theories, including Einstein gravity,
that are correct at some level but are not tree-unitary, this situation is
suggestive of new physics.
\item{(b)} When we come to $SU_q(3)$ we have found that the external algebra is
certainly not complete by itself, 
in the sense that a local theory along the lines possible for $SU_q(2)$ is no longer possible.

Both (a) and (b) suggest that the internal algebra should be included to
complete the theory, especially since we should expect that the external
theory describes only perturbative aspects of the full theory.  
A gauge invariant version of the internal
theory has been proposed in Ref. 2 at the level of $SU_q(2)$ and a similar
construction may be possible for $SU_q(3)$.

The construction of a local gauge theory derived from quantum groups has remained a challenge since their discovery, and there are now many papers in
this extensive literature dating from the early 1990's.  
Unfortunately these remain largely formal.

\vskip.5cm

\line{{\bf References.} \hfil}
\vskip.3cm

\item{1.} R. Finkelstein, Int. J. Math. Phys. A{\bf 18}, 627 (2003).
\item{2.} R. Finkelstein, hep-th/0206021 and Lett. Math. Phys. {\bf 62},
199 (2002).
\item{3.} J. Fuchs, {\it Affine Lie Algebras and Quantum Groups}.
\item{4.} S. Gasiorowicz, {\it Elementary Particle Physics}.
\item{5.} A. Gavrilek, Nucl. Phys. B, Proc. Suppl., Vol. 102-103, pp. 298-308.
\item{6.} J. M. Cornwall {\it et al.}, Phys. Rev. D{\bf 10}, 1145 (1974).

\end
\bye